\documentclass[aip,rsi,preprint,amsmath,amssymb]{revtex4-1}

\usepackage{textcomp}
\usepackage{graphicx}

\begin{document}




\keywords{dust dynamic, sheath, fast imaging, rescaled range analysis, automatic tracking.}


\title[Investigating transport of dust particles in plasmas]{Investigating transport of dust particles in plasmas}

\author{S. Bardin}
\email{sebastien.bardin@lpmi.uhp-nancy.fr}
      
\author{J-L. Brian\c{c}on}
\author{F. Brochard}

\affiliation{Institut Jean Lamour, U.M.R. CNRS 7198, Nancy-Universit\'e, Facult\'e des Sciences et Techniques, Boulevard des Aiguillettes, BP 70239, F-54506 Vandoeuvre-l\`es-Nancy Cedex,  France} 

\author{V. Martin}
\affiliation{CEA Cadarache, INRIA, Sophia Antipolis, France} 

\author{Y. Zayachuk}

\author{R. Hugon}
\author{J. Bougdira}

\affiliation{Institut Jean Lamour, U.M.R. CNRS 7198, Nancy-Universit\'e, Facult\'e des Sciences et Techniques, Boulevard des Aiguillettes, BP 70239, F-54506 Vandoeuvre-l\`es-Nancy Cedex,  France}            

\begin{abstract}

An algorithm has been developped, which makes it possible to automatically extract trajectories of a large number of particles from fast imaging data, allowing a statistical analysis of particles trajectories under various plasma conditions, a better understanding of their influence on plasma properties, and a better characterization of the plasma itself. In this contribution, we focus on results obtained in a radiofrequency parallel plate reactor, where a large amount of micron-sized carbon dust is produced in situ. The use of the rescaled range analysis (R/S analysis) applied to dust particles displacements allows decomposing dust dynamic on different time scales. It is shown that dust displacement is dominated by collisions on short time scales whereas long term behaviour is strongly influenced by large scale plasma fluctuations. 
  
\end{abstract}

\maketitle

\section{Introduction}

During the last decade dusty plasmas have attracted much attention because of their various applications, notably in plasma processing technologies. The mechanisms of generation and transport of such particles are still extensively investigated. It was shown that, either for individual dust grains or for large concentration of dust particles, dynamic and spatial distribution strongly depend on a variety of forces acting on them \cite{cava}. Depending on operating conditions, equilibrium between these forces varies and is responsible of dust cloud shape and dynamic.
The investigation of dust particles dynamic can allow obtaining some plasma parameters or dust particle characteristics such as their size \cite{wang,taka}. In many cases, oscillations are observed which result from a perturbation in the balance between forces acting on dust grains. An oscillator equation can usually be derived, showing that the oscillation amplitude and period are mainly function of particle mass, charge, and electric field \cite{zafi}. Then, working with calibrated dust grains, the latter can be used as probes to infer some plasma parameters such as the electric field \cite{basn}. On the contrary, the knowledge of plasma parameters makes it possible to estimate the mass and/or charge of dust grains by studying their trajectories. In both cases, the knowledge of dust particle response to a variation of the control parameter is needed. Such studies can be really time consuming and difficult to carry out, especially when particles are numerous and/or with complex trajectories. 
In all cases, it is necessary to have access either to a statistical analysis of a large number of dust particles trajectories, or to have some accurate dust trajectories on a long duration giving an estimation of the overall motion.
To perform such an analysis, an algorithm has been developed, which allows obtaining dust particles trajectories automatically, from camera measurements showing even a large concentration of particles (several thousands) and with a rather simple experimental set-up (e.g., no external light source is needed). In this contribution, this powerful diagnostic is used to evidence different types of dust response to sheath oscillations.
Paper is organized as follow: the description of the experimental setup, ultra fast camera diagnostic and operating parameters are given in sec.2. A brief description of main forces acting on dust particles in our plasma configuration is given in sec.3. Main observations concerning the effects of gas injection on dust displacements and collective dynamic are presented in sec.4. Two very different kinds of behavior are underlined in absence and presence of gas inlet, which can be explained by the use of the R/S analysis.

\section{Experimental setup}

For investigation of dust transport, macroscopic (tens of $\mu$m and larger) carbon dust particles were generated in capacitive coupled RF parallel plate reactor working at frequency 13.56 MHz, with upper electrode being biased and lower one grounded. Vacuum in reactor chamber of this device is created by means of rotary vane pump. Plasma was initiated in mix of argon and acetylene ($C_{2}H_{2}$). The motion of dust particles was recorded by an ultra fast camera. Besides direct detection and tracking of dust particles, camera observation allowed obtaining information about sheath dynamic, such as direct measurements of sheath luminescence intensity and thickness. Gases are continuously pumped out, simultaneously with gas injection. Mass flow meters control gas injection, which is set independently for each component of desired gas mixture. For experiments performed in this study it is necessary for dust particles to move approximately in single vertical plane. For this reason, electrodes used were rectangular, with length 7 cm, width 2 cm and thickness 0.7 cm. The key element of experimental setup is fast camera FASTCAM SA1.1 by Photron Ltd. Image recording is provided by a 12 bits monochrome CMOS sensor, with a spatial resolution one mega pixel up to 5400 fps. To improve resolution, external objective Nikon Micro-NIKKOR 105 mm was mounted. With this objective, a resolution of 20 $\mu$m per pixel was achieved.
Particles size depending on operating parameters, these ones must be well chosen to achieve satisfying conditions for our experiments. For this purpose, RF power ranges from 10 W to 20 W, and pressure of gas mixture (14 \% of $C_{2}H_{2}$ and 86 \% of Ar) is ~ 80 mTorr. This range of values allows obtaining dust particles with sufficient size to be observable by fast imaging camera. For a frame rate of 1000 fps, particle mean displacement between two frames is usually less than 1 pixel, which is well suited for tracking.
A new diagnostic, still under development, based on detection and tracking of dust particles, is used to obtain dust trajectories. It can operate in various experimental conditions (low temperature or fusion plasmas) and studies (statistical analysis or accurate individual trajectories). A complete description of the procedure is discussed in references \cite{zaya,ends}.
                                              
\section{Forces acting on spatial distribution and transport of dust particles}

Even in the simplest case (in the absence of magnetic field) of laboratory plasmas, such as the following performed experiments, dust particles in plasma volume are subject to the action of various forces which confine them in the plasma or drag them outside (see Fig. 2). Assuming negatively charged spherical particles, according to the literature \cite{robe}, the charge for an isolated dust grain is given by:
\begin{equation}
Q_{d} = 4\pi\varepsilon_{0}r_{d}V_{d}
\end{equation}
with $\epsilon_{0}$ the vacuum permeability, $r_{d}$ the radius of the dust particle and $V_{d}$ its electrostatic potential. 
All the forces in competition depend on the size of dust particles (15 $\mu$m or more in our case), either directly or through the electric charge and then can be distinguished in two classes: those independent of the charge (gravitational, neutral drag and thermophoretic forces, respectively $F_{g}, F_{n}, F_{th}$ in Fig.2 and those directly determined by the charge (electrostatic force $F_{el}$, ion drag force $F_{ion}$ and interaction between particles, which cannot be neglected in our case because of the high density of dust)\cite{cava}. 
In typical capacitive coupled RF discharges, like the one used in the following experiments, the most essential forces acting on micron-sized dust particles are ion drag force (which tends to confine them), electrostatic force and gravitational force (which tend to eject them from the plasma). Note that because of the low RF power (typically 10 W), temperature gradient in plasma chamber is small, and thus thermophoretic force can be neglected. The following section is dedicated to the study of dust particles motion under influence of sheath oscillations.

\section{Influence of sheath oscillations on dust transport.}

As already mentioned, normal operation of RF generator used in experiments implies continuous inlet and outlet of gas. RF generator can be operated in different regimes when gas pumping in and pumping out are switched off so that the amount of gas and pressure remain constant. Dust trajectories in presence and in absence of gas inlet exhibit strong differences, as illustrated in Fig.3. When there is no gas inlet, particles generally grow at a given place and then fall down due to gravity. In the opposite case, trajectories of lots of particles exhibit clear tendency for particles to move in the horizontal direction. This direction corresponds to the propagation direction of sheath waves, evidenced by camera observations as can be seen in Fig.4. As explained in reference \cite{zaya}, these oscillations are directly linked to potential fluctuations. By disturbing the force balance, these ones can clearly influence dust confinement and transport. This property was used to validate the reliability of the tracking algorithm by forcing dust oscillations \cite{zaya}. However, in this particular case, low frequency oscillations ($\sim$ 10 Hz) were mandatory to induce a collective response of dust to a forced potential perturbation. In general cases, the frequencies of spontaneous oscillations are much higher, of the order of 100-1000 Hz.
The fact that sheath oscillations are related to gas motion is proved by the observation, that the latter strongly influences the amplitude and frequency of sheath fluctuations. To make a comparison, frequency and amplitude parameters for two cases only differentiated by gaz injection are adduced in Table 1, with light intensity standard deviation being the parameter, characterizing fluctuation amplitude. 
Notably, the amplitude of sheath fluctuations differs in different regions of the sheath – namely below, above and from the side of the cathode – when gas is being pumped in, while when there is no gas inlet, deviation is much smaller, indicating that sheath fluctuations are weak, and have virtually the same magnitude all around the cathode.
In order to describe the transport of dust under the influence of a large variety of operation conditions, it is necessary to characterize the main feature of particles motion with a simple universal parameter. For that purpose, we propose to use the Hurst exponent $H$. The estimation of $H$ using the R/S analysis, is used to evidence any long-period dependance in the temporal behaviour of a given phenomenon \cite{hurst}. Such an analysis requires the knowledge of successive positions of single particles for a time scale statistically large compared to the characteristic evolution time of the trajectories. At the moment, the current procedure allows obtaining trajectories for about 1000 frames at the best, which is still not fully satisfactory. For that reason, only a rather preliminar analysis illustrating the benefits of this technique is presented hereafter. For a more accurate analysis, longer trajectories requiring further improvements of the tracking algorithm are necessary.
The value of $H$ is normalized to 1, which corresponds to a purely ballistic trajectory. $H = 0.5$ characterizes a brownian motion, and values between 0 and 0.5 indicate antipersistent phenomena. Fig.5 shows the Hurst exponents $H_{x}$ and $H_{y}$ respectively calculated in the directions perpendicular and parallel with respect to the cathode, for a single particule under the influence of spontaneous fluctuations seen in Fig. 4.  Whereas high values of $H_{x}$ and $H_{y}$ point out similar behaviours for short time delays ($\tau \leq 40$ frames), evidencing a classical diplacement between two collisions, a clear discrepancy appears for long time delays. The motion in the x direction, perpendicular to sheath oscillations, is in between a fractional Brownian motion and a ballistic one, whereas dynamic in the y direction is clearly antipersistent. This is a signature of the strong influence of the fluctuations on dust transport. These observations are confirmed in all cases where long enough trajectories are processed. It proves that the transport of dust, even the massive ones, is strongly impacted by fast fluctuations.

\section{Conclusion}

In order to investigate dust transport in a laboratory plasma, fast camera imaging was performed. Dust trajectories were extracted with an automatic procedure and the advantages of the R/S analysis for obtaining main features of dust trajectories was demonstrated. This technique will be used to study dust transport in different discharges conditions in order to determine if plasma fluctuations can be used to control dust transport. 

\begin{acknowledgements}
  
This work, supported by the European Community under the contract of
Association between EURATOM, CEA, and the French Research Federation for
fusion studies, was carried out within the framework of the European
Fusion Development Agreement. The views and opinions expressed herein do
not necessarily reflect those of the European Commision. Financial
support was received from the French National Research Agency
(contract ANR-08-JCJC-0068-01).

\end{acknowledgements}

\hfil
\begin{table}
\caption{Characteristics of light intensity fluctuations in different regions of the sheath.}
\label{tab:1}
\begin{tabular}{@{}lllll@{}}
\hline
 &  & Below the cathode & Above the cathode & From the side of the cathode \\
\hline
Gaz inlet present & Frequency (Hz) & 129.3 & 107.7 & 110.7 \\
 & Standard deviation & 2.7 & 19.8 & 5.4 \\
Gas inlet absent & Frequency (Hz) & 163.3 & 167.8 & 163.7 \\
 & Standard deviation & 1.4 & 1.7 & 1.9 \\
\hline
\end{tabular}
\end{table}
\hfil

\begin{figure}
\begin{minipage}{80mm}
\includegraphics[width=5cm, height=4cm]{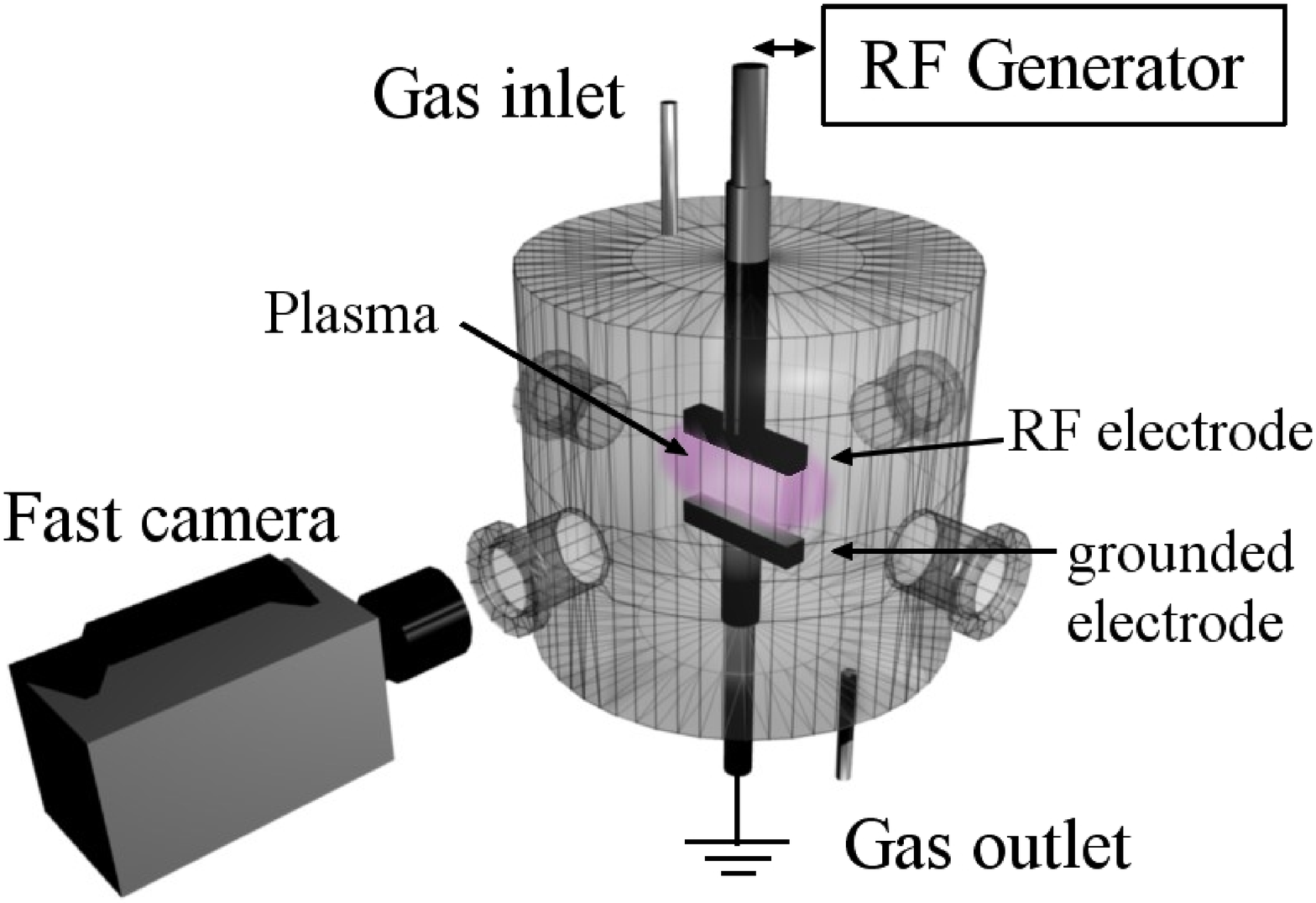}
\caption{General scheme of experimental setup.}
\label{fig:1}
\end{minipage}
\hfil
\begin{minipage}{70mm}
\begin{center}
\includegraphics[width=5cm,height=4cm]{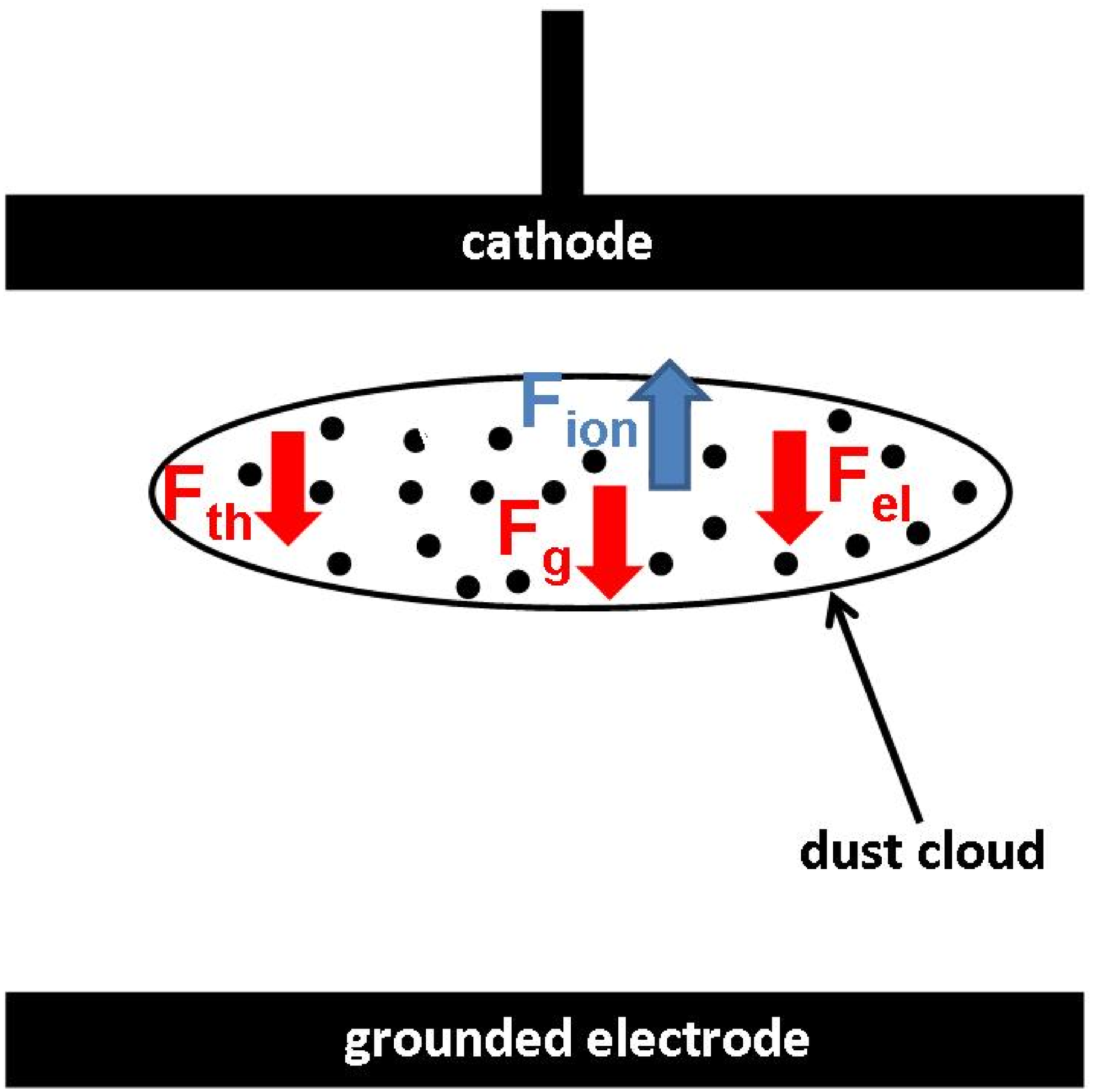}
\caption{Representation of the different forces acting on an isolated dust grain.}
\label{fig:2}
\end{center}
\end{minipage}
\end{figure}

\hfil

\begin{figure}
\includegraphics[width=7.5cm,height=3.5cm]{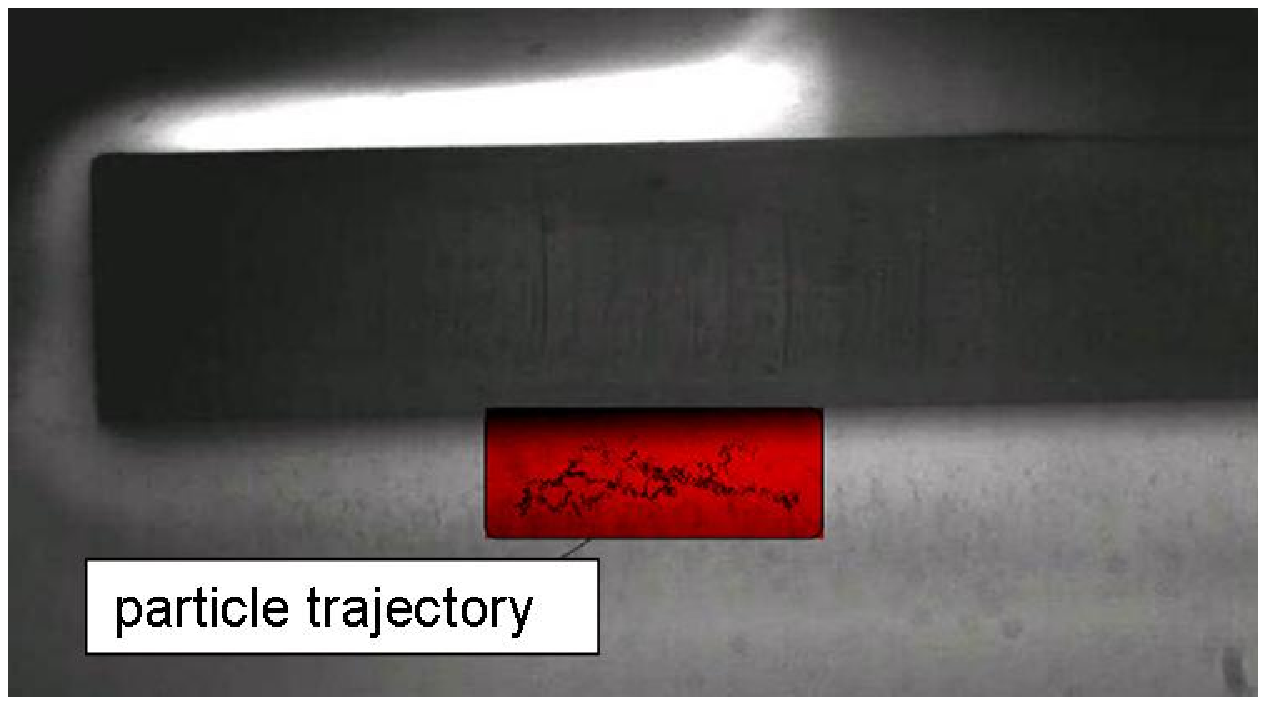}~a)
\hfil
\includegraphics[width=7.5cm,height=3.5cm]{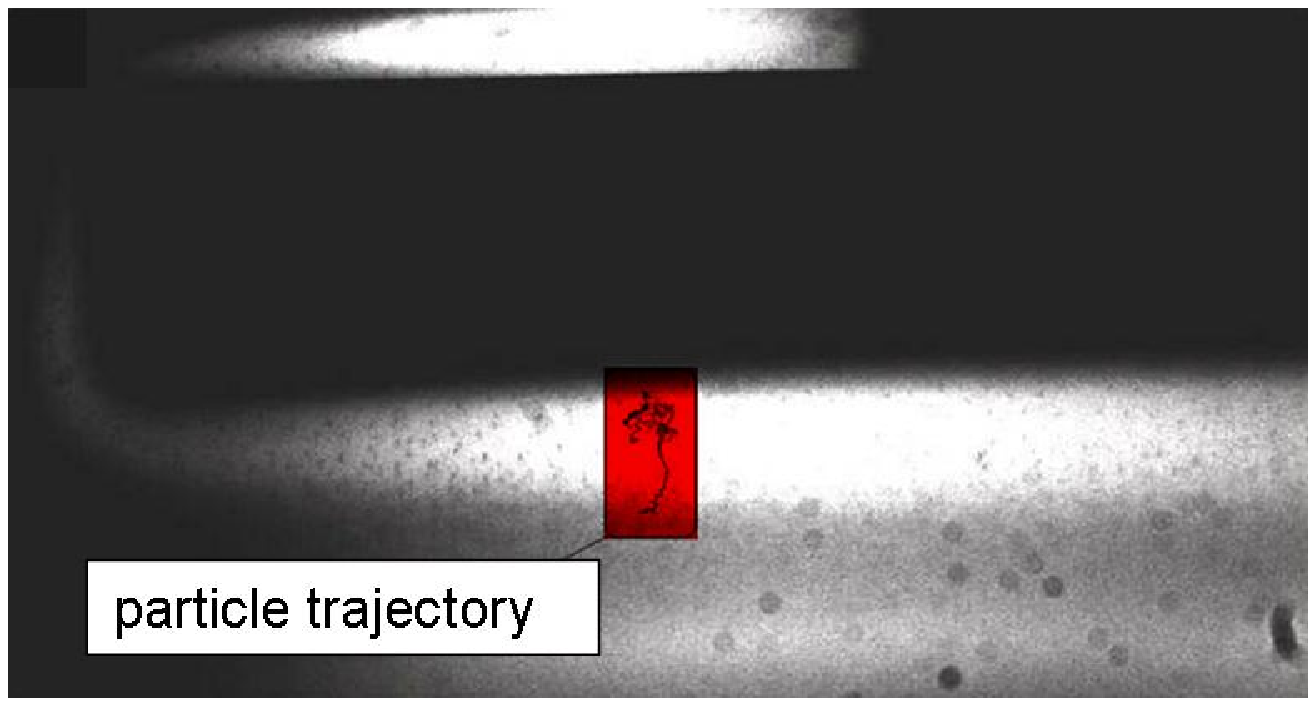}~b)
\caption{Comparison of particle trajectories \textbf{a)} in presence and \textbf{b)} in absence of gas inlet.}
\label{fig:3}
\end{figure}
\hfil

\begin{figure}
\begin{minipage}{60mm}
\includegraphics[width=7cm, height=4cm]{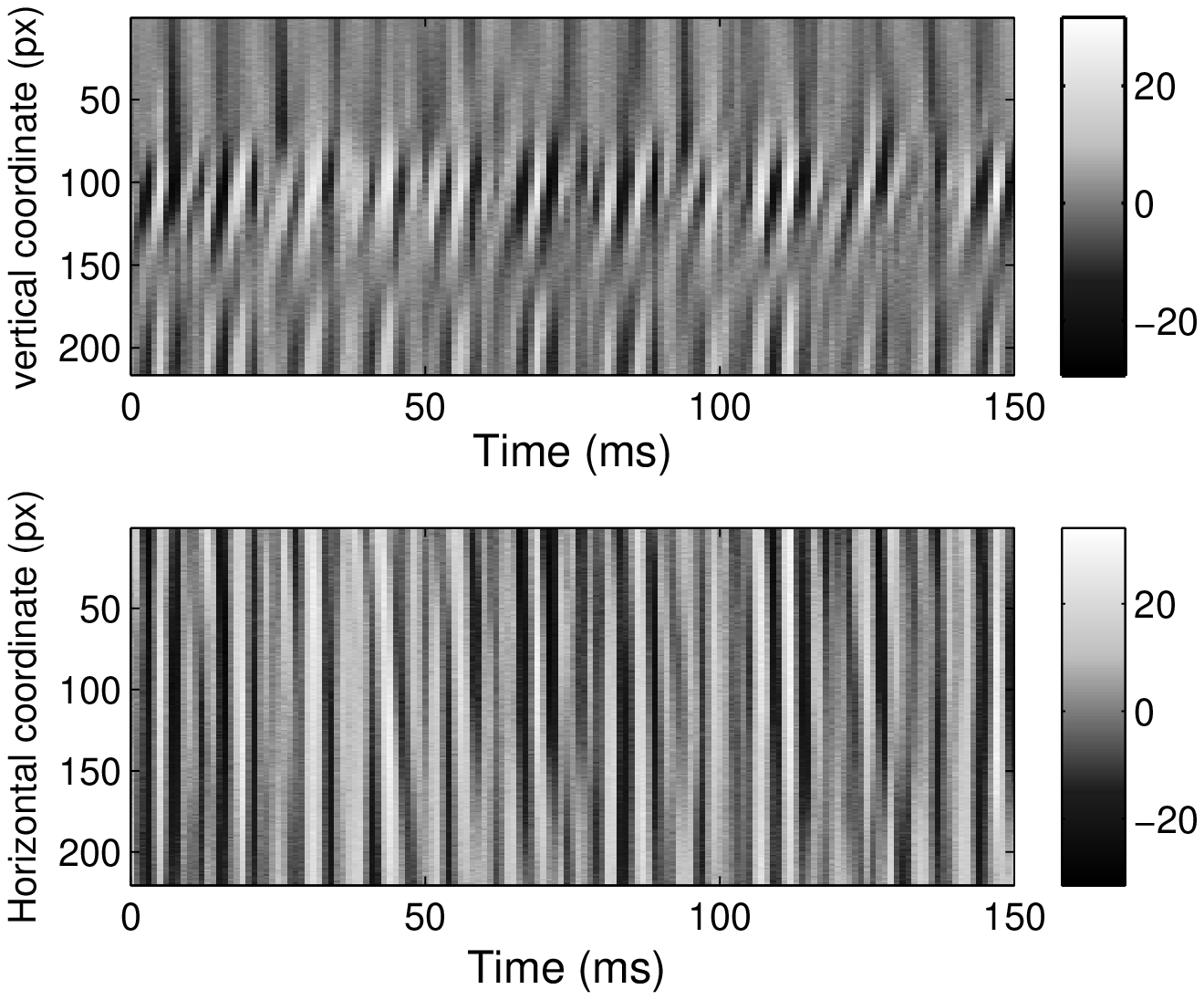}
\caption{Temporal dependance of sheath light fluctuations along vertical (top) and horizontal (bottom) axes, evidencing vertical propagation.}
\label{fig:4}
\end{minipage}
\hfil
\begin{minipage}{80mm}
\begin{center}
\includegraphics[width=8cm,height=4cm]{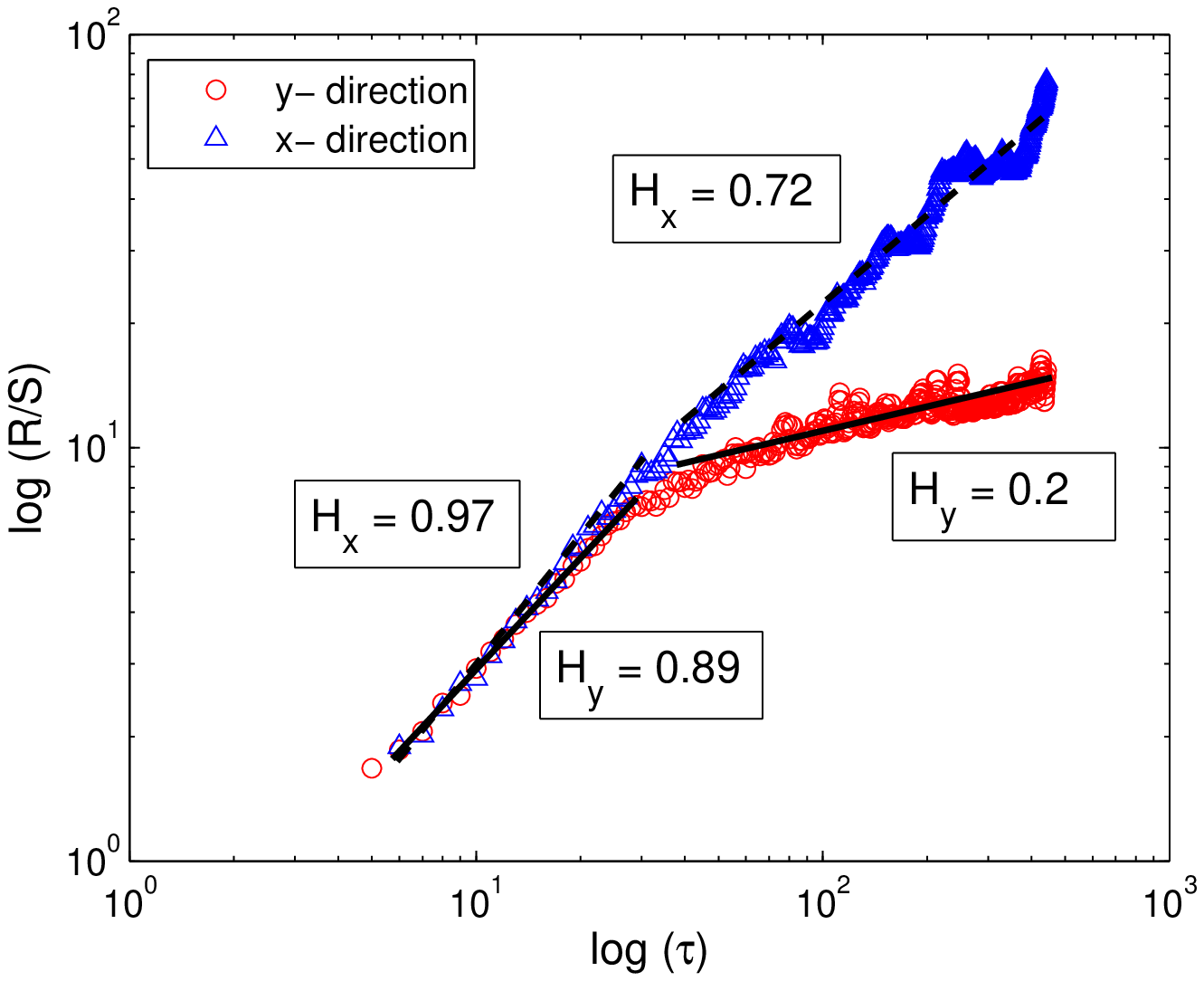}
\caption{Hurst exponents corresponding to vertical ($H_{y}$) and horizontal ($H_{x}$) displacements of a single dust particle for different time scales.}
\label{fig:5}
\end{center}
\end{minipage}
\end{figure}

\end{document}